\documentclass[aps,prc,twocolumn,amsmath,amssymb,superscriptaddress,showpacs,showkeys,preprintnumbers,nofootinbib]{revtex4-2}
\usepackage{dcolumn}
\usepackage{graphicx,color}
\usepackage{multirow}
\usepackage{textcomp}
\usepackage{physics}
\usepackage{dsfont}
\usepackage{relsize}
\usepackage{appendix}

\usepackage{tikz}	
\begin{document}
	\newcommand {\nc} {\newcommand}
	\nc {\beq} {\begin{eqnarray}}
	\nc {\eeq} {\nonumber \end{eqnarray}}
	\nc {\eeqn}[1] {\label {#1} \end{eqnarray}}
\nc {\eol} {\nonumber \\}
\nc {\eoln}[1] {\label {#1} \\}
\nc {\ve} [1] {\mbox{\boldmath $#1$}}
\nc {\ves} [1] {\mbox{\boldmath ${\scriptstyle #1}$}}
\nc {\mrm} [1] {\mathrm{#1}}
\nc {\half} {\mbox{$\frac{1}{2}$}}
\nc {\thal} {\mbox{$\frac{3}{2}$}}
\nc {\fial} {\mbox{$\frac{5}{2}$}}
\nc {\la} {\mbox{$\langle$}}
\nc {\ra} {\mbox{$\rangle$}}
\nc {\etal} {\emph{et al.}}
\nc {\eq} [1] {(\ref{#1})}
\nc {\Eq} [1] {Eq.~(\ref{#1})}
\nc {\Refc} [2] {Refs.~\cite[#1]{#2}}
\nc {\Sec} [1] {Sec.~\ref{#1}}
\nc {\chap} [1] {Chapter~\ref{#1}}
\nc {\anx} [1] {Appendix~\ref{#1}}
\nc {\tbl} [1] {Table~\ref{#1}}
\nc {\Fig} [1] {Fig.~\ref{#1}}
\nc {\ex} [1] {$^{#1}$}
\nc {\Sch} {Schr\"odinger }
\nc {\flim} [2] {\mathop{\longrightarrow}\limits_{{#1}\rightarrow{#2}}}
\nc {\textdegr}{$^{\circ}$}
\nc {\inred} [1]{\textcolor{red}{#1}}
\nc {\inblue} [1]{\textcolor{blue}{#1}}
\nc {\IR} [1]{\textcolor{red}{#1}}
\nc {\IB} [1]{\textcolor{blue}{#1}}
\nc {\IG} [1]{\textcolor{green}{#1}}
\nc{\pderiv}[2]{\cfrac{\partial #1}{\partial #2}}
\nc{\deriv}[2]{\cfrac{d#1}{d#2}}

\nc {\bit} {\begin{itemize}}
	\nc {\eit} {\end{itemize}}

\title{The role of the likelihood for elastic scattering uncertainty quantification}
\author{C.~D.~Pruitt}
\email{pruitt9@llnl.gov}
\affiliation{Lawrence Livermore National Laboratory, P.O. Box 808, L-414, Livermore, California 94551, USA}
\author{A.~E.~Lovell}
\email{lovell@lanl.gov}
\affiliation{Theoretical Division, Los Alamos National Laboratory, Los Alamos, New Mexico, 87545, USA}
\author{C.~Hebborn}
\affiliation{Facility for Rare Isotope Beams, Michigan State University, East Lansing, Michigan 48824, USA}
\affiliation{Department of Physics and Astronomy, Michigan State University, East Lansing, Michigan 48824, USA}
\author{F.~M.~Nunes}
\affiliation{Facility for Rare Isotope Beams, Michigan State University, East Lansing, Michigan 48824, USA}
\affiliation{Department of Physics and Astronomy, Michigan State University, East Lansing, Michigan 48824, USA}

\date{\today}
\preprint{LLNL-JRNL-860637}
\begin{abstract}
\begin{description}
\item[Background] Analyses of elastic scattering with the optical model (OMP) are widely used in nuclear reactions.

\item[Purpose] Previous work compared a traditional frequentist approach and a Bayesian approach to quantify uncertainties in the OMP. In this study, we revisit this comparison and consider the role of the likelihood used in the analysis.

\item[Method] We compare the Levenberg-Marquardt algorithm for $\chi^{2}$ minimization with Markov Chain Monte Carlo sampling to obtain parameter posteriors. Following previous work, we consider how results are affected when $\chi^{2}$/N is used for the likelihood function, N being the number of data points, to account for possible correlations in the model and underestimation of the error in the data.

\item[Results] We analyze a simple linear model and then move to OMP analysis of elastic angular distributions using a) a 5-parameter model and b) a 6-parameter model. In the linear model, the frequentist and Bayesian approaches yield consistent optima and uncertainty estimates. The same is qualitatively true for the 5-parameter OMP analysis. For the 6-parameter OMP analysis, the parameter posterior is no longer well-approximated by a Gaussian and a covariance-based frequentist prediction becomes unreliable. In all cases, when the Bayesian approach uses $\chi^{2}$/N in the likelihood, uncertainties increase by $\sqrt{N}$.

\item[Conclusions] When the parameter posterior is near-Gaussian and the same likelihood is used, the frequentist and Bayesian approaches recover consistent parameter uncertainty estimates. If the parameter posterior has significant higher moments, the covariance-only frequentist approach becomes unreliable and the Bayesian approach should be used. Empirical coverage can serve as an important internal check for uncertainty estimation, providing red flags for uncertainty analyses.

\end{description}

\end{abstract}

\maketitle
%

\section{Introduction}
\label{sec:intro}

Nuclear reactions and the reaction theory used to interpret them are crucial in a variety of applications of nuclear science.  Because the interpretation of  measurements depend on the theoretical models, significant effort has been put into developing methodology for the quantification of their uncertainties.  Most of the work has been focused on quantifying parametric uncertainties \cite{Lovell2015,Lovell2017,King2018,King2019,Catacora2019,Catacora2021,Whitehead2022,Surer2022,Catacora2023,Hebborn2023,Hebborn2023PRL} that results from fitting models to data with experimentally reported uncertainties. 
Over the past several years, the paradigm for optimization and uncertainty quantification (UQ) has evolved. Previously, the standard approach involved least-square optimization (i.e., minimizing $\chi^2$) and propagation of uncertainty through estimation of a covariance matrix (referred to here as {\it frequentist approach}). In the last few years, Bayesian methods have become more prominent. The shift is due both to the philosophical approach underpinning Bayesian methods and because these methods provide information about the full parameter space, as sampled through direct Monte Carlo methods \cite{Trotta2008}. 

In reaction theory, this progression has been performed systematically. A decade ago uncertainties on reaction observables, such as cross sections were typically estimated by qualitative comparison of observables generated by two candidate phenomenological potentials or two few-body approximations, e.g.~\cite{Lovell2015}. Then, a framework for propagating uncertainties from parametric covariance matrices that included theoretical correlations was developed~\cite{Lovell2017}. In elastic scattering, these theoretical correlations arise from  the construction of the cross section as a sum over Legendre polynomials~\cite{ReactionsBook}. When taken into account in the optimization procedure, both the optimum and the resulting uncertainty intervals change. Similarly, were the full correlations on the experimental data--which are rarely reported--accounted for in the likelihood function used for optimization, the resulting optimum and uncertainty estimates would change\footnote{These types of differences have been shown in e.g.~\cite{Neudecker2020} but have not necessarily been studied in the context of the two optimizations frameworks that we are discussing here.}.

A comparison between frequentist and the Bayesian approaches was performed in Ref.~\cite{King2019} for the optical model, a widely used model in nuclear reactions. Parameters of the model were fit to the elastic scattering  angular distribution for a given projectile-target combination at a given beam energy and the resulting parametric uncertainties were propagated to cross sections for one-neutron transfer reactions. The frequentist approach involved least-squares fitting using {\sc sfresco}, a widely used reaction code that includes least-squares optimization tools \cite{fresco}. The Bayesian method involved Markov Chain Monte Carlo sampling with the Metropolis algorithm \cite{quilt-r}. In that work, it was found that uncertainties produced with the Bayesian implementation were larger than those using the standard frequentist approach.

There are at least two aspects of that study that contribute to this surprising finding. The first has to do with degeneracies in parameter space: a single elastic scattering measurement--one target-projectile combination at one incident energy--is sufficient only to constrain the real and imaginary volume integrals of the potential used. Because the potential included both imaginary surface and imaginary volume Woods-Saxon terms, each with a depth, radius, and diffuseness, fitting at a single scattering energy was insufficient to break the degeneracy between imaginary terms.
The authors addressed this in their frequentist treatment by fixing several parameters during the optimization, which therefore did not enter into the calculation of the parameter covariance matrix, see e.g.~\cite{Lovell2021,Lovell2017}. In the Bayesian approach, the authors instead applied priors that enforced physicality of the parameters, and allowed all parameters to vary. 
The second contributing factor is the choice of the likelihood function, which was not consistent in the two approaches.
Since the intent of that work was to connect with the standard procedures in the field, in the frequentist approach, the standard $\chi^2$ minimization was used (see e.g. \cite{Lovell2017,King2018}). However, for reasons discussed below, the Bayesian approach used $\chi^2/N$ for the likelihood instead, where $N$ is the number of points in a data set. In the present work, we will revisit the frequentist/Bayesian comparison considering first a simple linear problem and various likelihood functions and then we will add complexity to the problem to illustrate how the comparison may yield different results, including where a Bayesian treatment provides advantage for scattering problems.


The outline of this manuscript is as follows.  In Sec.~\ref{sec:stats}, we detail the statistical considerations of this paper, including Bayes' theorem, the role of the likelihood in the specific case of elastic scattering, and how correlations in the experimental data impact the optimization and uncertainty quantification.  In Sec.~\ref{sec:toyModel}, we compare the frequentist and Bayesian optimizations in detail, performing a one-to-one comparison with a toy model; the toy model is followed up with an example of typical optical model application for the interpretation of elastic scattering experiments in Sec.~\ref{sec:realCase}.  Finally, we draw conclusions in Sec.~\ref{sec:conclusions}. 

\section{Statistical considerations}
\label{sec:stats}


In this section, we summarize the key elements in the frequentist and Bayesian approaches when applied to an optical model (see \cite{Lovell2017,lovell2018} for more details). In an optical model analysis, elastic scattering angular distribution data $\frac{d\sigma}{d\Omega}$, depending on scattering angle $\theta$, are used to calibrate the model. For simplicity in notation, we will use $\sigma(\theta)$ to represent the angular distribution $\frac{d\sigma}{d\Omega}$.

{\it Frequentist approach:} The frequentist likelihood function minimizes the squared difference between experimental data and the corresponding optical model prediction for $n$ model parameters, $x$:

\begin{equation}
\label{eqn:chi2}
{\chi^2} =  \sum \limits _{i=1} ^N  \frac{[\sigma_\mathrm{exp} (\theta_i)- \sigma _\mathrm{th} (\theta_i, x)]^2}{[\Delta \sigma _\mathrm{exp}(\theta_i)]^2},
\end{equation} 
where $N$ is the number of data points, $\sigma_\mathrm{exp}(\theta_i)$ is the experimental cross section at angle $\theta_{i}$, $\sigma_\mathrm{th}(\theta_{i},x)$ is the theoretical cross sections at angle $\theta_{i}$,
and $\Delta \sigma _\mathrm{exp}(\theta_i)$ is the experimental uncertainty on the measurement, $\sigma _\mathrm{exp}(\theta_i)$. Equation \eqref{eqn:chi2} is simply a weighted least-squares approximation that assumes that the data covariance matrix contains experimental errors (on the diagonal) and nothing else, as is typical in previous scattering-data analyses. Optimization may be performed by a variety of classical algorithms, including
gradient descent, simulated annealing, or Levenberg-Marquardt \cite{Levenberg, Marquardt}. 

Once an optimum is reached, uncertainties about that optimum can be obtained by calculating moments of the parameter distribution. Typically, only the second moment
-- the covariance -- is assessed, which can be done by numerical estimation and inversion of the Hessian matrix. By truncating the moment expansion at the second moment, we assume $x$ is well-represented by a multivariate Gaussian distribution about the optimum. Uncertainties on the observables can be calculated by sampling from this multivariate Gaussian and propagating sampled parameter sets to the observable of interest. 
Inherently, uncertainty quantification from the $\chi^2$ minimization assumes a unimodal distribution of the parameters. If the distribution has significant higher moments or other modes that are not included in uncertainty estimation, it can lead to unrealistic uncertainty estimates in observables.

Finally, the analyst may choose to rescale the parameter uncertainties by invoking a ``goodness-of-fit" metric, $\chi^{2}/d$, where $d$ is the number of independent degrees of freedom $d=N-n$. (While the concept of an ``independent degree of freedom'' is fraught in a non-linear model, in the present analysis we will assume that each parameter is independent to simplify comparison of the uncertainty quantification approaches we investigate). This degree-of-freedom normalization guarantees that, if the data uncertainties can be relaxed to be \textit{relative} rather than absolute, the parameter optimum will yield a perfect fit of the data, up to the data uncertainty: $\chi^{2}/d = 1$.

{\it Bayesian approach:} Markov Chain Monte Carlo, the Bayesian method considered here, samples parameter space via multiple independent walkers. 
Bayes' theorem~\cite{Phillips2021} is: 
\begin{equation}
    p(H|D,M)=\frac{p(D|H,M)p(H|M)}{p(D|M)}.
\end{equation}
Here $p(H|M)$ is the prior distribution, the information known about the parameter distribution $H$ of a specific model $M$ before seeing the data $D$; $P(D|H,M)$ is the likelihood, which assesses the probability of observing data $D$, assuming model $M$ and a given parameter set; $p(D|M)$ is the Bayesian evidence; and $p(H|D,M)$ is the posterior distribution providing updated information about the parameter values after information from the data are included. Because the Bayesian evidence typically includes an integral over the entire parameter space of the model, this integral is often numerically challenging or even impossible to compute. Thus, implementations of the Bayesian framework often use Markov-Chain Monte Carlo (MCMC), which does not require computing the Bayesian evidence, to draw posterior-distribution samples. Note that if the Bayesian evidence $p(D|M)$ is not computed, then the Bayesian approach can no longer provide an absolute probability of a given parameter vector, only the relative probability between samples, as the MCMC posterior is no longer normalized.

In MCMC analysis, the multivariate weighted-least-squares likelihood is \cite{KDUQ}:
\begin{equation}
p(D|H,M) = \frac{1}{\sqrt{(2\pi)^{n} \prod\limits_{i}^{N} \Delta\sigma_{exp}(\theta_{i})^{2}}}\exp \left [ -\chi^2/2 \right ].
\label{eq:Lchi2}
\end{equation}

Here $n$ is the dimensionality of the model parameters, and $N$ is the number of data points, with $\chi^{2}$ the same as in Eq.~\eqref{eqn:chi2}. If only relative likelihoods are required, the normalizing factor may be dropped:
\begin{equation}
    p(D|H,M) = \exp \left [-\chi^{2}/2 \right].
\end{equation}

\textit{Correlations:} In principle, both experimental and theoretical correlations exist and should be taken into account explicitly in the likelihood function:
\begin{equation}
\label{eqn:corrChi2}
\chi^2 = [\sigma_\mathrm{exp} (\theta_i)- \sigma _\mathrm{th} (\theta_i, x)] \mathbb{C} ^{-1} [\sigma_\mathrm{exp} (\theta_i)- \sigma _\mathrm{th} (\theta_i, x)] ^T
\end{equation}
where $\mathbb{C}$ is a combination of the experimental and theoretical covariance matrices. This correlation-cognizant approach -- generalized-least-squares -- is not broadly used in scattering analyses because $\mathbb{C}$ is not known in most scattering problems, so the weighted-least-squares approach Eq.~\eqref{eqn:chi2} is the most commonly employed. It represents the extreme case of no correlations between angles (uncorrelated experimental uncertainties and no theory uncertainties or correlations).

\vspace{0.5cm}

In previous studies in direct reaction theory~\cite{Lovell2015,Lovell2017,King2018,King2019,Catacora2019,Catacora2021,Whitehead2022,Surer2022,Catacora2023,Hebborn2023,Hebborn2023PRL}, a different likelihood function was used for Bayesian analyses only:
\begin{equation}
p(D|H,M) = \exp \left [ -\chi^2/(2N) \right ],
\label{eq:Lchi2N}
\end{equation}
where $\chi^2$ is defined as in Eq. (\ref{eqn:chi2}). The original motivation for including the $1/N$ factor was the acknowledgement of model correlations \cite{Lovell2017}.
Assuming the experimental errors are accurate (they include not only statistical but all systematic errors), the $1/N$ weighting ensures that when two data sets $\sigma_{th}(\theta)$ are included in an optimization, one with twice as many angles than the other, both contribute equally to the optimization, based on the uncertainties that were assigned by the experimental group that performed the measurement (see~\cite{TemplateIntro} for a more extended discussion). The $1/N$ factor was used to represent an assumption of perfect correlation between angles for a given experimental angular distribution -- that is, if the cross section is measured at one angle, the entire angular distribution can be extracted. While this is true for, e.g., Rutherford scattering, realistic correlations for optical-potential-driven scattering lie somewhere between the uncorrelated weighted-least-squares case [Eq. \eqref{eqn:chi2}] and the ``perfectly correlated" $\chi^{2}/N$ case. Alternatively, rescaling the likelihood by $1/N$ is equivalent to increasing each uncertainty $\Delta_\mathrm{exp}(\theta_i)$ by a factor of $\sqrt{N}$. In other words, it is an assertion that the experimentally reported uncertainties are too small by a factor of $\sqrt{N}$. The impact of rescaling the likelihood by $N$ is the topic of Sec. \ref{sec:toyModel} and Sec. \ref{sec:realCase}.

\section{Results for a linear toy model}
\label{sec:toyModel}

With the above statistical framework in mind, we present several examples relevant to recent UQ studies in nuclear physics. First, we show that for linear models calibrated to linear data with Gaussian noise, frequentist and Bayesian treatments will recover the \textit{same} parametric uncertainty estimates, and we demonstrate that uncertainty estimates recovered from either method scale with the likelihood function used in the calibration.

Consider the following toy problem: an analyst is presented with noisy experimental data and wishes to train a linear model to estimate the relationship present in the data. The true function used to create these data is:
\begin{equation}
    \begin{split}
        y & = 2 - 0.5x + \delta \\
        \delta & \sim \mathcal{N}(0, 0.01),
    \end{split}
\end{equation}
where $\mathcal{N}$ indicates a normal-distributed random variable with the mean 0 and variance 0.01. The experimental independent variable $x$ has $N=25$ points in a grid ranging from -1 to 1.

We consider four possible approaches for calibrating the model parameters and assessing parameter uncertainties. In all approaches the same linear model was used:
\begin{equation}
    y = mx + b,
\end{equation}
with the parameters $m$ and $b$ initialized to zero.

\bit
\item 
The first calibration (which we refer to as LM-$\chi^2$) uses a standard least-squares likelihood function Eq.~(\ref{eq:Lchi2})
and the Levenberg-Marquardt (LM) algorithm \cite{Levenberg, Marquardt} for parameter optimization. This algorithm blends the Gauss-Newton method with gradient descent and is representative of frequentist optimization approaches discussed in recent publications \cite{Lovell2015,Lovell2017,King2018,King2019}. In this work, we used the LM implementation provided by the \textsc{lmfit} python3 module, version 1.2.2 \cite{lmfit}. 

\item
The second calibration (which we refer to as MCMC-$\chi^2$) uses a standard least-squares likelihood function Eq.~(\ref{eq:Lchi2}) and Markov-Chain Monte Carlo (MCMC) for parameter optimization. Here, we used the python3 MCMC library \textsc{emcee} \cite{emcee} with the default Goodman-Weare stretch-move proposal distribution and flat priors spanning the range of physical plausibility, in keeping with \cite{King2019}. We note that our results were largely insensitive to MCMC hyperparameter and proposal distribution changes, provided that at least 5000 steps were taken following a burn-in period. Walkers were initialized with a small amount of Gaussian noise $\epsilon\approx10^{-4}$ about zero, enabling the stretch-move proposal step. 

\item
For the third calibration (which we refer to as LM-$\chi^2/\text{d}$), we take the results of the first approach and rescale the parameter uncertainty estimates $\Delta x$ post-facto by $\chi^2/\text{d}$, where d$=N-n$ are the degrees-of-freedom.
This goodness-of-fit rescaling approach was used in \cite{King2019} to perform frequentist calculations. 

\item
For the fourth and last calibration (which we refer to as MCMC-$\chi^2/N$) we use MCMC as in the second calibration above, but with a modified likelihood function as in Eq. \eqref{eq:Lchi2N}.
This approach was used in Ref.~\cite{King2019} as the example of Bayesian methodology.
\eit

\begin{figure}
    \centering
    \includegraphics[width=0.5\textwidth]{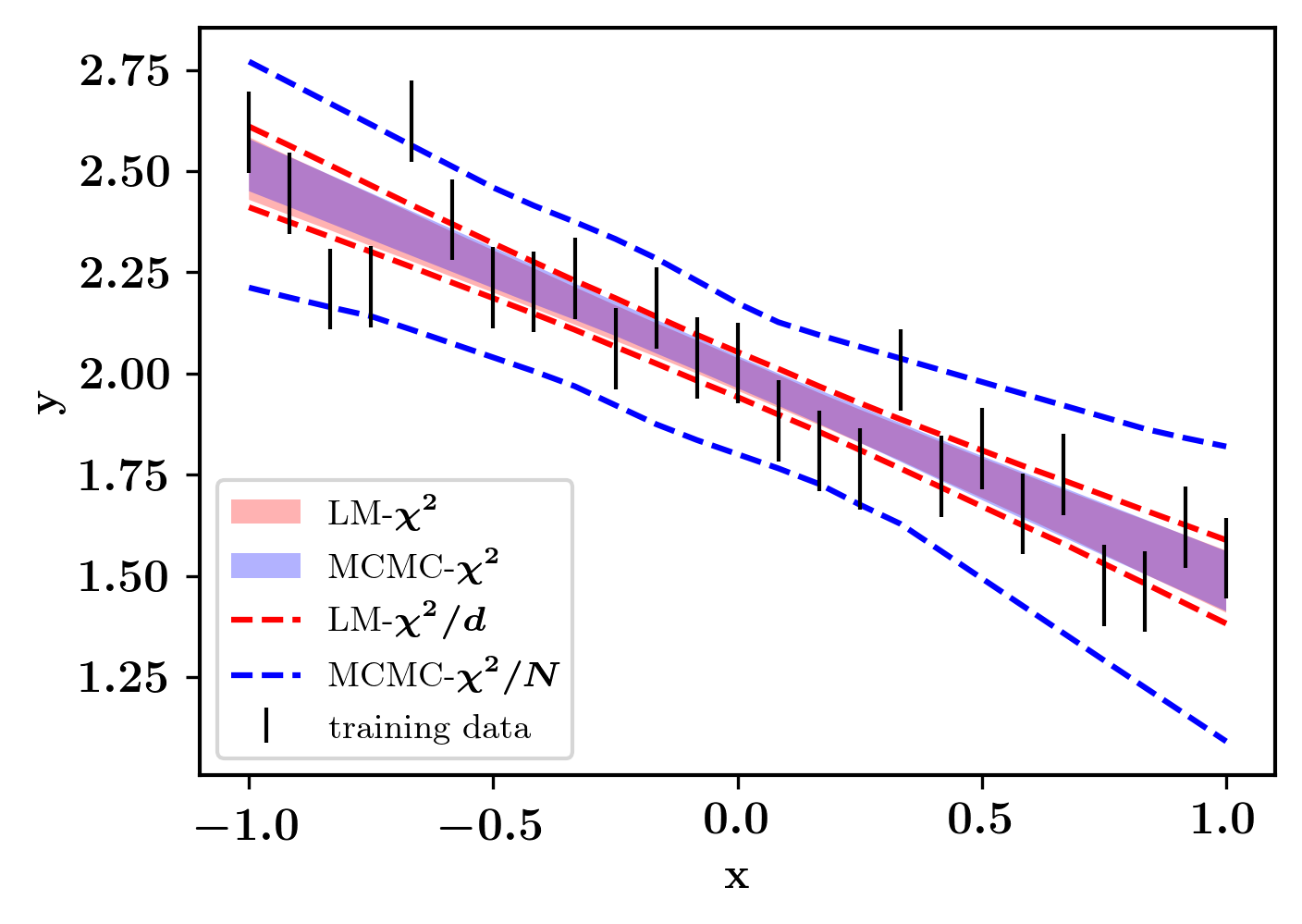}
    \caption{Uncertainty estimates from four different calibration approaches are compared for a linear toy model. ``LM-$\chi^2$" refers to the Levenberg-Marquardt approach, using the nominal ($-\chi^{2}/2$) log-likelihood function. ``MCMC-$\chi^2$" refers to Markov-Chain Monte Carlo, using the nominal log-likelihood function. ``LM-$\chi^{2}/\text{d}$" refers to the Levenberg-Marquardt approach using the nominal log-likelihood function, but with the final uncertainty estimates rescaled by the reduced chi-square at the optimum. ``MCMC-$\chi^2/N$" refers to Markov-Chain Monte Carlo, but rescaling the nominal log-likelihood function by $1/N$, where $N$ is the number of experimental data points. For each model, the band represents the 95\% uncertainty interval. For the training data, the line length represents $\pm1\sigma$ experimental error.}
    \label{fig:linearModel}
\end{figure}

Figure \ref{fig:linearModel} 
shows the results from each of these calibration approaches on our toy linear model. All models recover the true parameters used to generate the data, but uncertainty estimates differ. We first compare the first and second method (blue and red shaded regions), which have the same likelihood but a different optimization method. LM-$\chi^2$ and  MCMC-$\chi^2$ recover consistent uncertainty intervals, with slight differences due to numerical truncation in the Hessian matrix estimate (LM-$\chi^2$) and statistical noise from the finite number of Monte Carlo samples (MCMC-$\chi^2$). Our finding is that for this linear case, the results are insensitive to the choice of optimization approach: MCMC and LM return statistically indistinguishable uncertainty estimates.

Next we consider the influence of modifications to the standard likelihood. The third method LM-$\chi^2/\text{d}$ recovers a slightly larger, 
though still consistent, uncertainty interval, a consequence of the goodness-of-fit metric, $\chi^2/d$, being slightly larger than 1. That the goodness-of-fit is close to unity is an indicator that the model form and experimental data span a consistent underlying distribution, guaranteed in this case by the toy problem construction. Most importantly, the fourth method MCMC-$\chi^2/N$ yields an uncertainty estimate that is $\sqrt{N} \approx 4.5$ times larger than the other distributions. This is because the likelihood function has been ``softened" by $N=25$, equivalent to increasing all experimental uncertainties by $\sqrt{N}$. The consistency of the frequentist and Bayesian, non-informative-prior approaches when using the \textit{same} likelihood function, and the inconsistency when using \textit{different} likelihood functions, is a general feature for linear models with an arbitrary number of data points or parameters. However, the same may not be not true for nonlinear models depending on the initialization conditions, as the likelihood surface may no longer be convex everywhere and multiple minima may be possible. We now turn to a non-linear model more relevant to nuclear reactions.
 
\section{Results for the optical model}
\label{sec:realCase}

To test a nonlinear example relevant to reaction theory, we revisit the optical model calibration study of~Ref.~\cite{King2019} using the four approaches discussed in Sec. \ref{sec:toyModel}.
In this section we present results for  \textsuperscript{90}Zr(n,n)\textsuperscript{90}Zr angular distributions at 10 MeV, using real data from Ref.~\cite{Wang1990} and including all $N=36$ points in the angular distribution. In addition, we also performed calculations for other cases considered in Ref.~\cite{King2019}, namely \textsuperscript{48}Ca(p,p) data and \textsuperscript{208}Pb(p,p) data at 16 MeV, and will comment on those following our analysis of the Zr data.

\subsection{Five-parameter optical model}
\label{sec:bg5}
As in Ref.~\cite{King2019}, we consider a simplified Becchetti-Greenlees (BG) optical model potential \cite{BG}, fitted to single-nucleon differential elastic scattering data.
As explained in the Introduction, because of degeneracy between the imaginary volume and imaginary surface potential terms, calibrating an optical potential with a single elastic scattering data set usually requires fixing potential parameters to lift the degeneracy. In Ref.~\cite{King2019}, the authors fixed the imaginary volume depth, radius, and diffuseness parameters when applying frequentist methodology. For our first example, we will fix these parameters and also the imaginary surface diffuseness, yielding a 5-parameter simplified Becchetti-Greenlees model. Thus only two potential terms are included in the optimization: the real volume term and the imaginary surface term. We allow the three parameters of the real volume potential to vary ($V$, $r$, $a$) as well as the depth and radius of the surface term, ($W_s$, $r_s$). 
This 5 parameter model is effective at reproducing the scattering data for both the LM and the MCMC calibration methods -- that is, $\chi^2/\text{d} \approx 1$ at the optimum. To compute scattering cross sections, we use the optical-potential library TOMFOOL \cite{KDUQ}.

\begin{figure}
    \centering
    \includegraphics[width=0.48\textwidth]{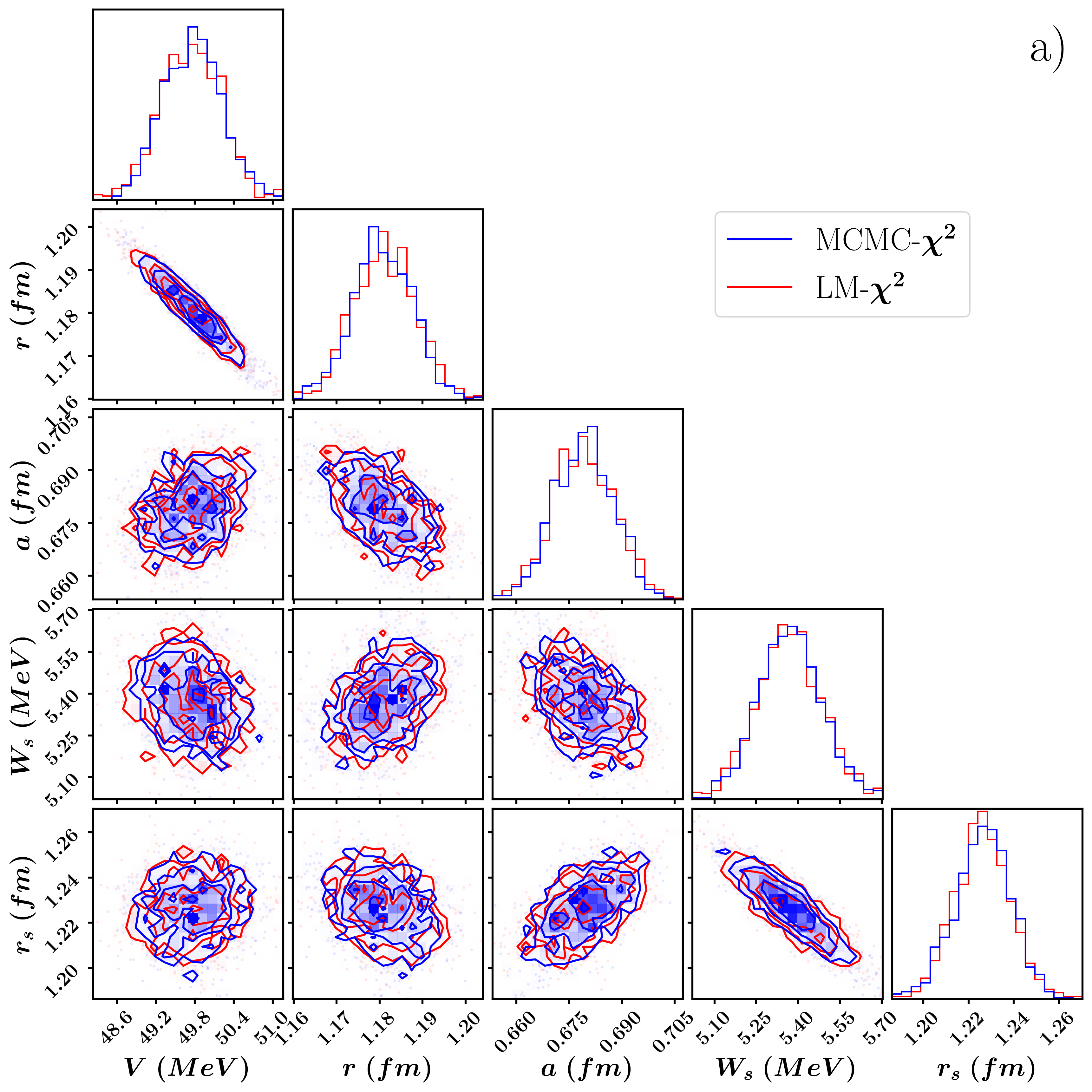}
    \includegraphics[width=0.48\textwidth]{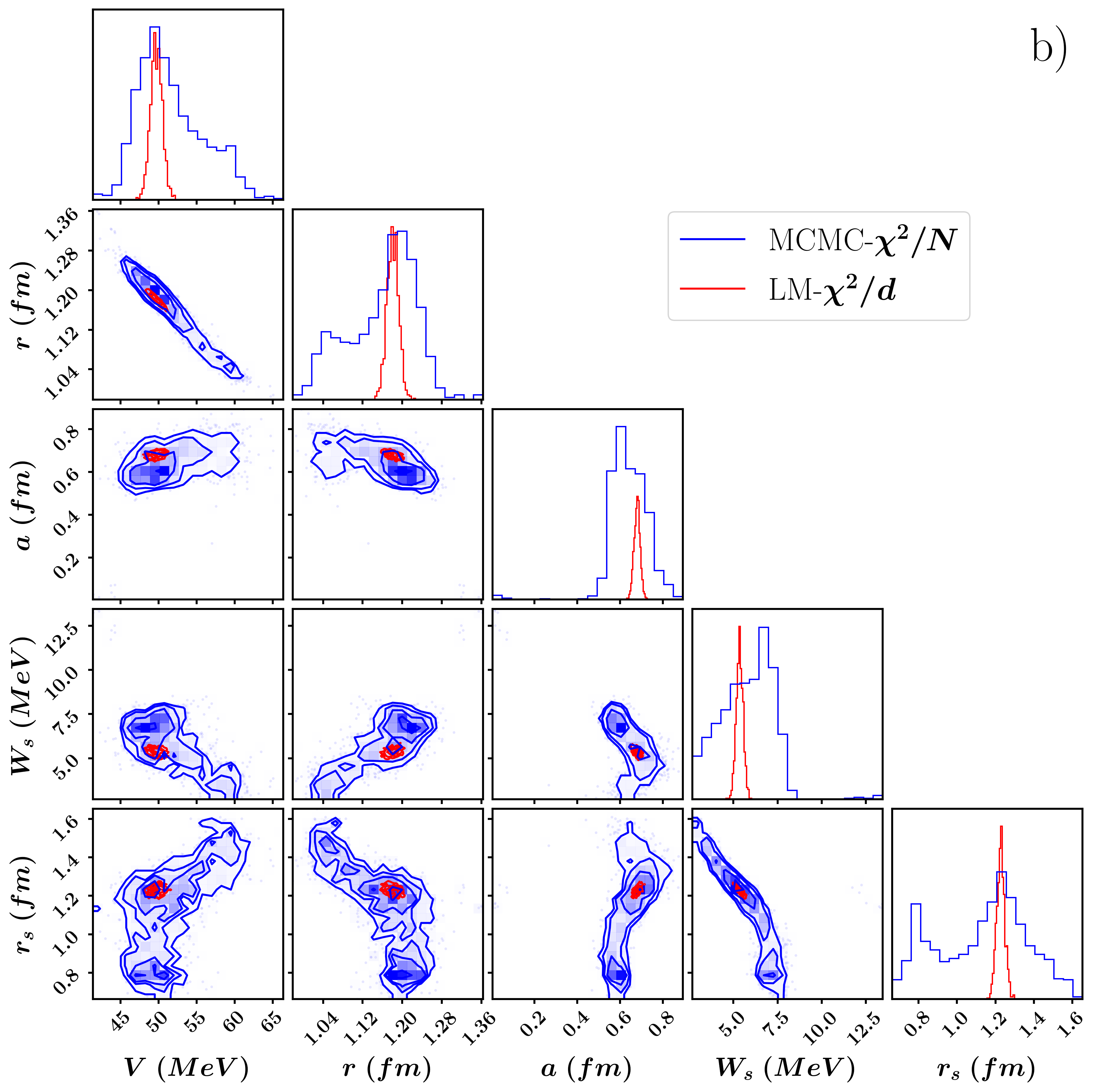}
    \caption{Corner plots \cite{corner} showing parameter distribution estimates from four calibration approaches applied to a simplified 5-parameter Becchetti-Greenlees optical model. Panel~a) shows results for the unscaled approaches ``LM-$\chi^2$'' and ``MCMC-$\chi^2$''. Panel b) shows results for the goodness-of-fit adjusted ``LM-$\chi^{2}/\text{d}$'' and for the likelihood-rescaled ``MCMC-$\chi^2/N$''.}
    \label{fig:BGToyModel_5param_corner}
\end{figure}

\begin{figure}
    \centering
    \includegraphics[width=0.48\textwidth]{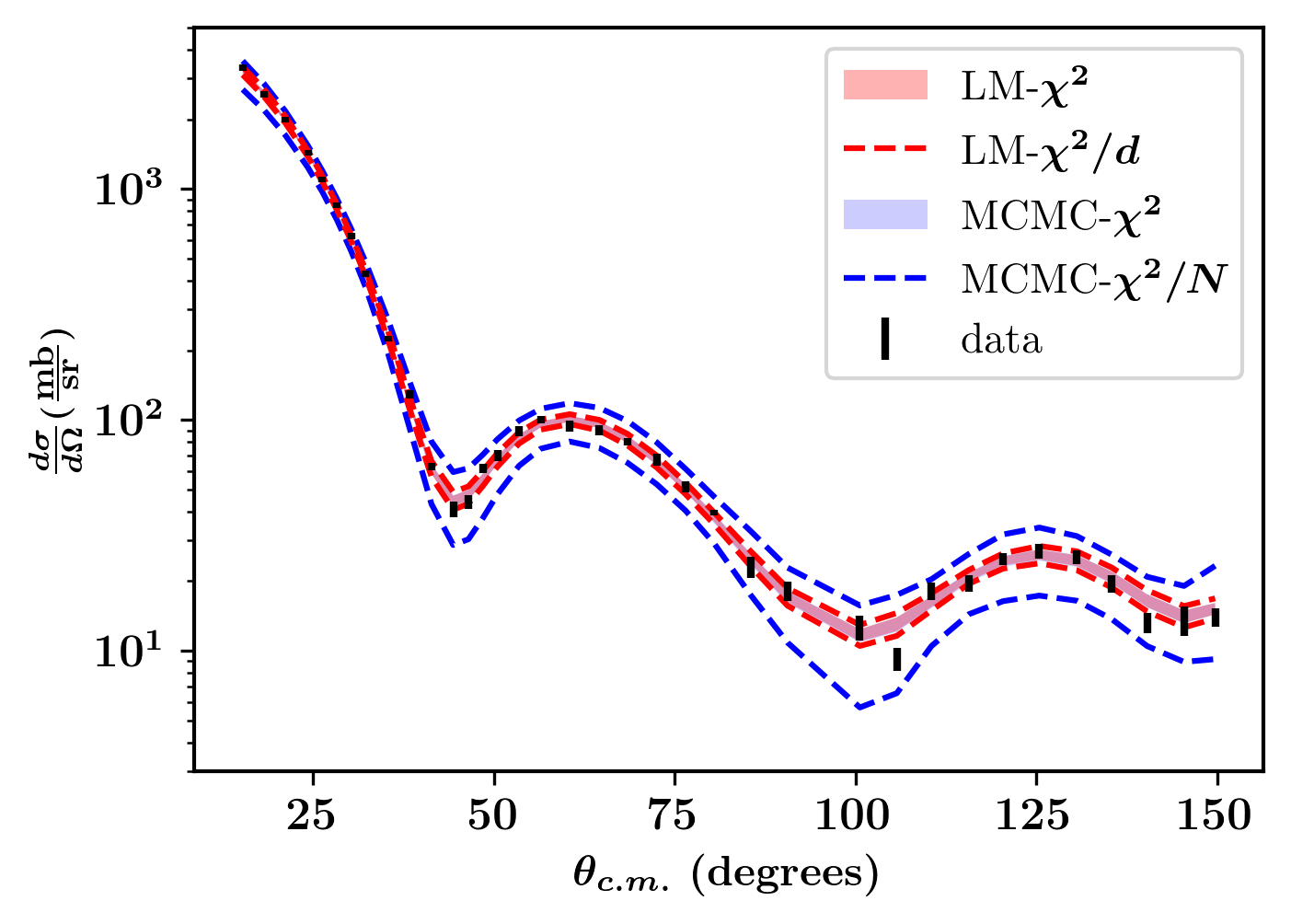}
    \caption{Cross section uncertainty estimates from four different calibration approaches for a simplified 5-parameter Becchetti-Greenlees optical model. Label convention is the same as in Fig. \ref{fig:BGToyModel_5param_corner}.}
    \label{fig:BGToyModel_5param}
\end{figure}

The results for the same four calibration approaches introduced earlier are presented in Figs.~\ref{fig:BGToyModel_5param_corner} and \ref{fig:BGToyModel_5param}.
The posterior distributions for LM-$\chi^2$ and MCMC-$\chi^2$,  shown in the diagonal of corner plot Fig.~\ref{fig:BGToyModel_5param_corner}(a), indicate that LM recovers the same parameter optimum and marginal uncertainty estimates as MCMC when the same likelihood is considered and the same set of parameters varied. Correlations between depth and radius parameters are consistent with the expectation that it is the volume integral of the potential, rather than the potential parameters themselves, that are constrained by data. In Fig.~\ref{fig:BGToyModel_5param_corner}(b) we show a corner plot comparing calibration approaches 3 and 4. Calibration approach 3 (LM-$\chi^2/d$) produces uncertainty estimates that are consistent with approaches 1 and 2 in panel a), a consequence of the fact that this five-parameter model is able to accurately reproduce these data and the ``goodness-of-fit" rescaling is small. The fourth approach, MCMC-$\chi^2/N$, produces uncertainty estimates that are several times larger than the other approaches.

To compare with the training data, we compute the $95$\% uncertainty intervals for the elastic angular distributions (Fig. \ref{fig:BGToyModel_5param}). As noted already, the red and blue shaded regions are fully consistent, even though this model is non-linear. The much-larger uncertainty for approach 4 is consistent with the results in \cite{King2019}, where MCMC uncertainties were found to be always larger than those obtained with the frequentist approach. As in the toy linear model presented above, the inconsistencies between approach 4 and the other approaches is due to the additional scaling factor in the likelihood function, not to improved fidelity of MCMC sampling.

\subsection{Six-parameter optical model}
\label{sec:bg6}

In certain non-linear problems, Bayesian calibration \textit{does} have an important functional advantage over uncertainty estimation via covariance -- but only when the likelihood surface around the parameter optimum has significant moments beyond the covariance. To demonstrate this, we modify the previous example to a six-parameter Becchetti-Greenlees model by allowing the imaginary surface diffuseness to vary.

\begin{figure}
    \centering
    \includegraphics[width=0.48\textwidth]{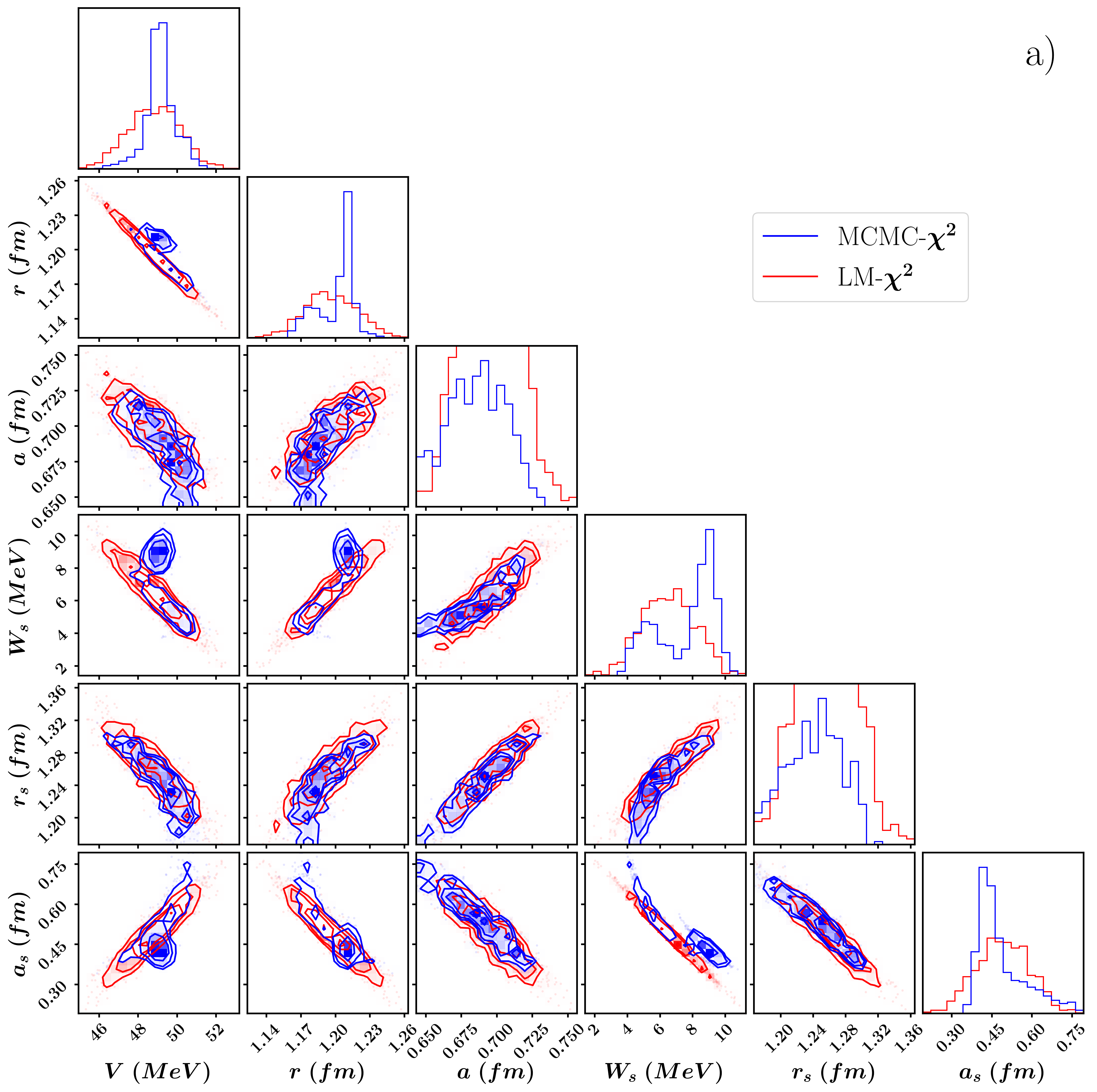}
    \includegraphics[width=0.48\textwidth]{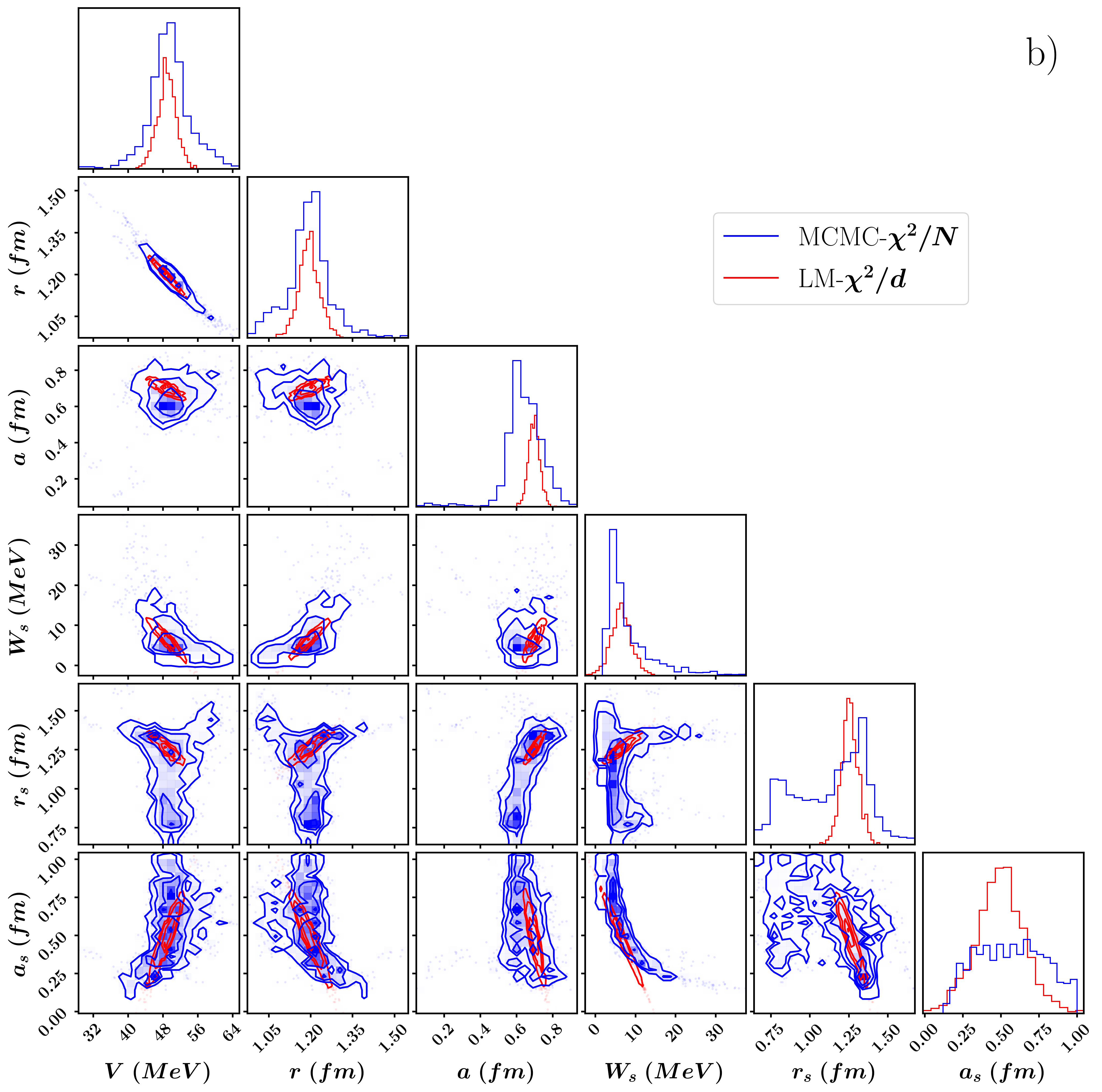}
    \caption{Corner plots \cite{corner} showing parameter distribution estimates from four calibration approaches applied to a simplified 6-parameter Becchetti-Greenlees optical model. Label convention is the same as in Fig. \ref{fig:BGToyModel_5param_corner}.}
    \label{fig:BGToyModel_6param_corner}
\end{figure}

\begin{figure}
    \centering
    \includegraphics[width=0.48\textwidth]{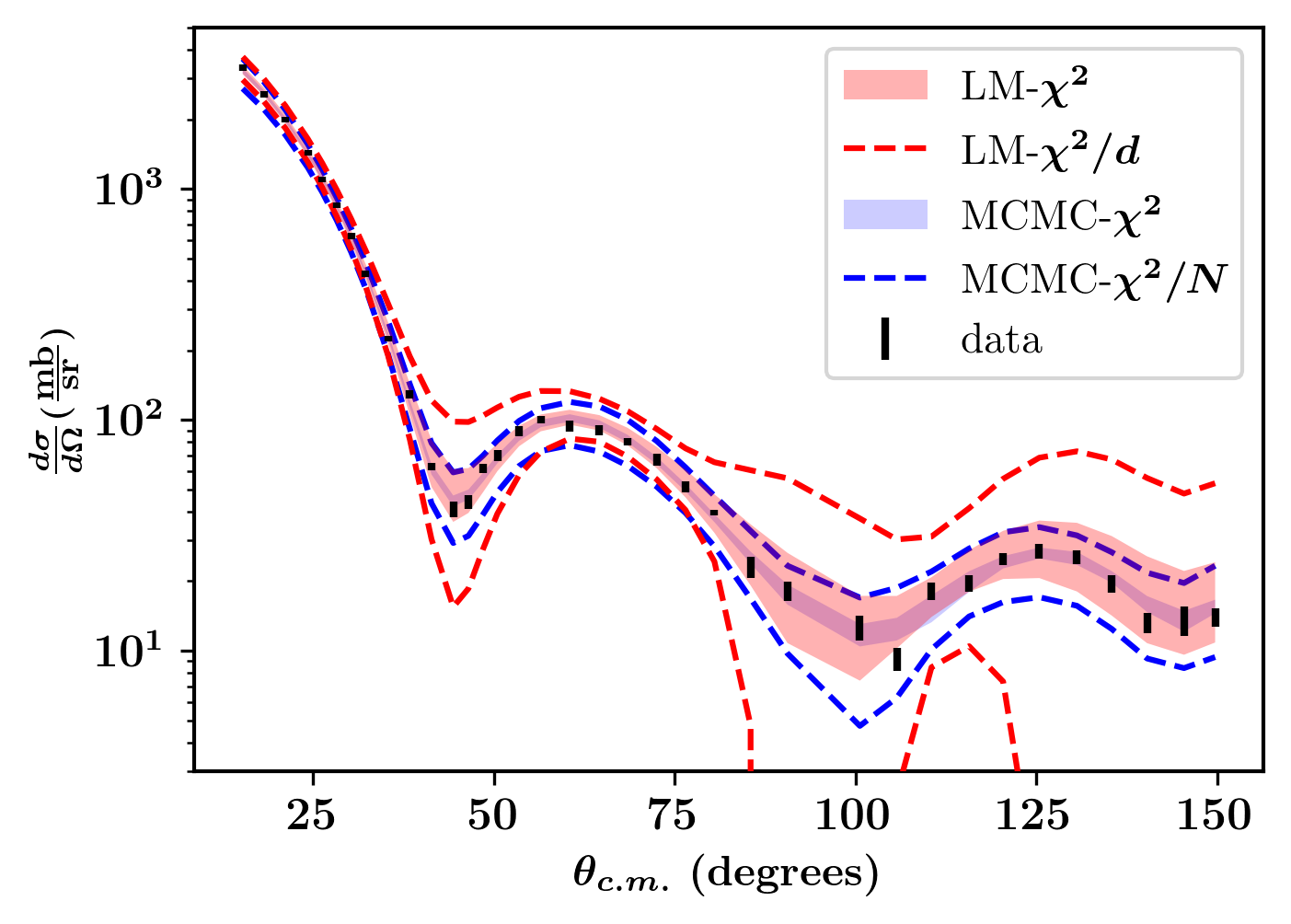}
    \caption{Cross section uncertainty estimates from four different calibration approaches for a simplified 6-parameter Becchetti-Greenlees optical model. Label convention is the same as in Fig. \ref{fig:BGToyModel_5param_corner}.}
    \label{fig:BGToyModel_6param}
\end{figure}

Figure \ref{fig:BGToyModel_6param_corner}(a) shows the parameter posteriors and the correlations for approaches 1 and 2 applied to this six-parameter model, and Fig. \ref{fig:BGToyModel_6param_corner}(b) shows the corresponding results for approaches 3 and 4. The Bayesian model reveals that, in the vicinity of the optimum, the newly enabled imaginary surface diffuseness and the imaginary surface depth have a complex, non-linear relationship (see blue banana shapes in Fig. \ref{fig:BGToyModel_6param_corner}(a)). In calculating the Hessian matrix to estimate uncertainties, the frequentist approach approximates this curved posterior distribution as Gaussian. This is a poor approximation that leads to an overestimate of the cross section uncertainty (red band in Fig. \ref{fig:BGToyModel_6param}).
If additional, imaginary-volume parameters are relaxed in the model, the degeneracy becomes so severe that the Hessian matrix cannot be numerically estimated.
Figure \ref{fig:BGToyModel_6param} shows how LM-$\chi^2/d$ can predict inappropriately large uncertainty estimates if the posterior distribution deviates from Gaussian. In this case, LM-$\chi^2$ and LM-$\chi^2/d$ require making assumptions that are not fulfilled by the parameter posteriors, so they will yield distorted parameter uncertainty estimates. As before, the blue band and the dashed blue lines continue to disagree by a similar amount as in Sec.~\ref{sec:bg5}, approximately $\sqrt{N}$, due to the rescaling of the likelihood function. The mismatch seen in Ref.~\cite{King2019} are due to the different scaling factors used in the frequentist vs. Bayesian likelihood, compounded by the use of a different number of model parameters when comparing the frequentist and Bayesian approaches. Given that each of the four calibration approaches yields a different uncertainty estimate in the six-parameter case, how should we assess their plausibility? To answer this question, we turn to the empirical coverage for each approach.

\subsection{Empirical coverages in the  optical model}
\label{sec:empcov}

\begin{figure}
    \centering
    \includegraphics[width=0.5\textwidth]{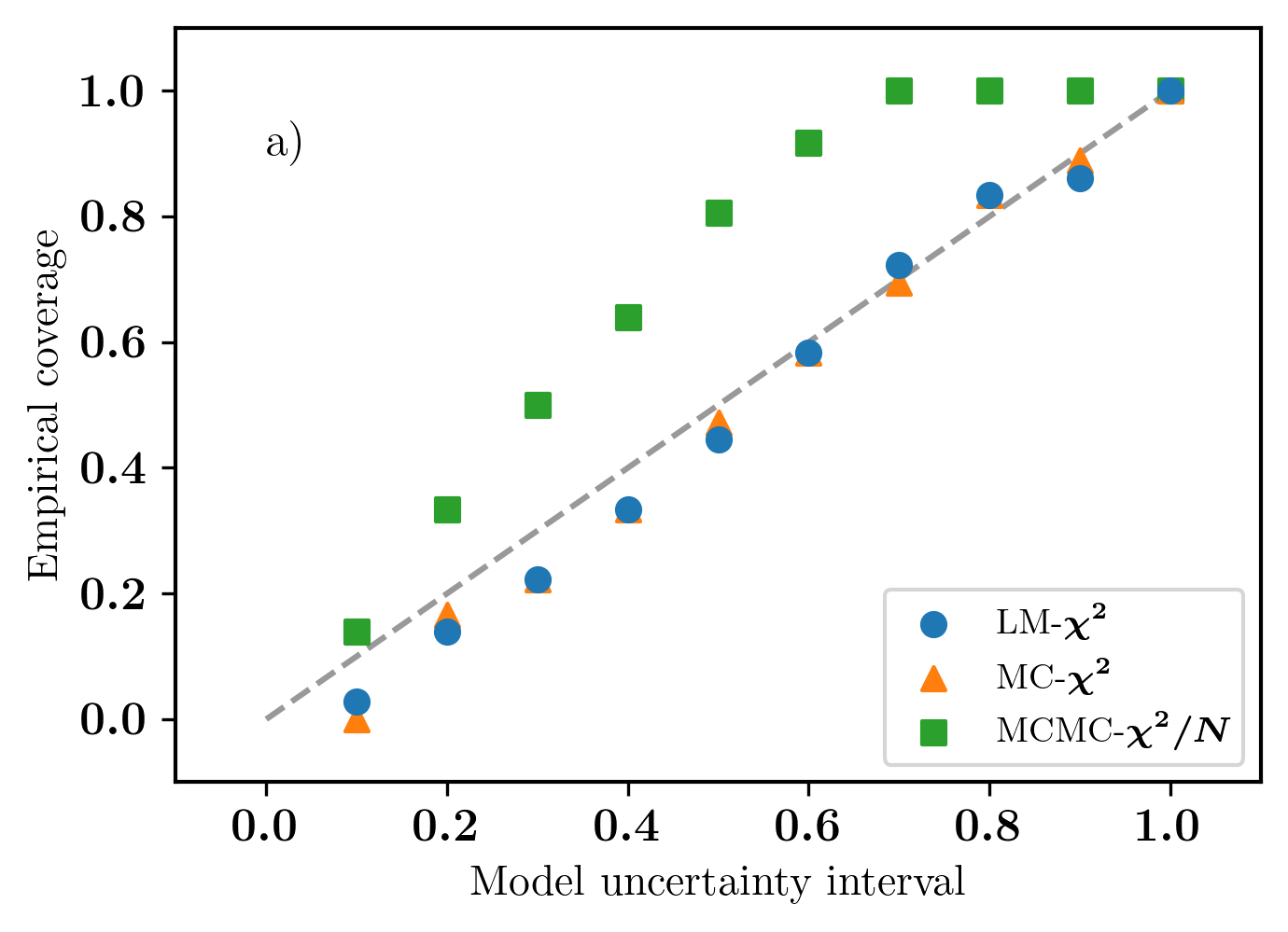}
    \includegraphics[width=0.5\textwidth]{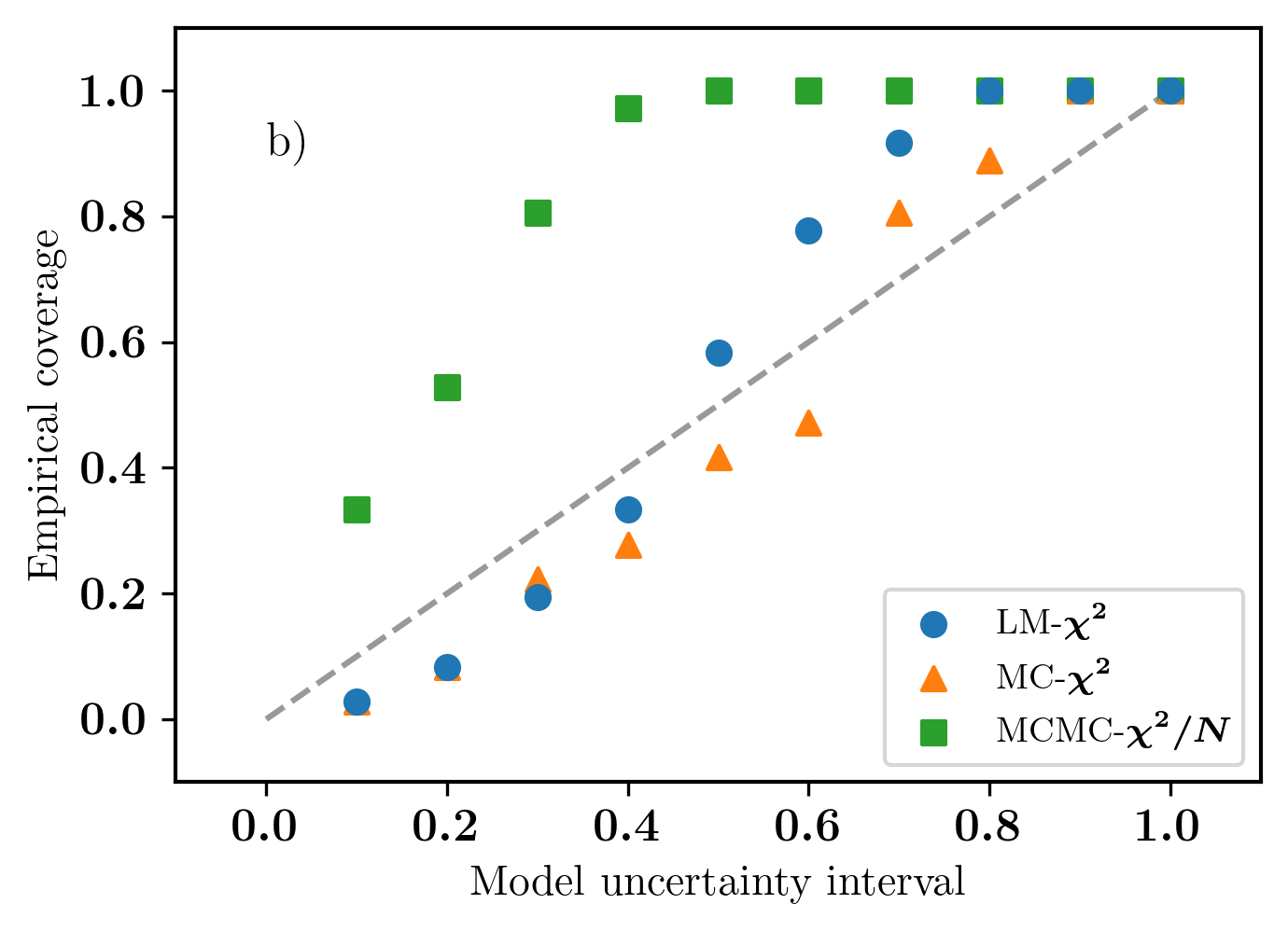}
    \caption{Empirical coverages from three calibration approaches are compared. Panel (a) shows results for the 5-parameters BG fit and panel (b) show results for the 6-parameters BG fit.}
    \label{fig:empcov}
\end{figure}

A sanity check for any method is that the predicted uncertainty interval spans a fraction of the underlying training data consistent with the uncertainty. For example, for a $68$\% uncertainty interval to be meaningful, it should span roughly $68$\% of the data. Empirical coverage diagrams formalize this idea by quantifying the fraction of experimental data that fall within the estimated model uncertainty intervals. Fig.~\ref{fig:empcov} shows empirical coverages for approaches 1, 2, and 4, with results for the 5-parameter model in panel a) and the 6-parameter model in panel b). Ideal empirical coverage is represented by the gray dashed diagonal.

We first consider the simpler 5-parameter BG calibration (Fig.~\ref{fig:empcov}(a)), a case where the parameter distributions are well-approximated by a multivariate Gaussian such that LM and MCMC produce near-identical results. As expected, the empirical coverage obtained for LM-$\chi^2$ and MCMC-$\chi^2$ (blue circles and orange triangles, respectively) are the same, allowing for small statistical uncertainty due to Monte-Carlo sampling in the MCMC result. Most importantly, they align with the diagonal, indicating that $1/N$ rescaling leads to overestimation of uncertainty (the green squares lie well above the diagonal).

Finally, we consider the 6-parameters BG fit (Fig.~\ref{fig:empcov}(b)). 
Again, the empirical coverage indicates that the uncertainties derived from MCMC-$\chi^2/N$ are overestimated no matter the uncertainty interval. When no normalization is added, LM-$\chi^{2}$ and MCMC-$\chi^2$, coverages are slightly below the diagonal ideal at the lowest uncertainty intervals and slightly above at the highest uncertainty intervals. Such deviations can indicate that one or more of the assumptions required for the statistical treatment -- that there are no outliers, no model defects, no off-diagonal terms in the data covariances -- may not be true. However, small deviations from the diagonal as in panel a) should be expected as an artifact of having a small number ($N$=36) of data points with which to populate the empirical coverage. 

We note that in calculating these empirical coverages, the experimental uncertainty for each point is added to the model's parametric uncertainty in quadrature (contrary to what was done in \cite{King2019}, where only the parametric uncertainty was considered for the empirical coverage estimate). Because the experimental uncertainties are small compared to the parametric uncertainty, this difference does not affect the results at a qualitative level; if the approach from Ref.~\cite{King2019} is used instead, the MCMC-$\chi^2$ result remains near the diagonal, while the MCMC-$\chi^2/N$ result sits well above.

In the optical potential analysis thus far, we have considered only the neutron scattering data on $^{90}$Zr from Ref.~\cite{King2019}. In Ref.~\cite{King2019}, analyses were conducted on several data sets spanning multiple targets, including proton and neutron scattering on $^{48}$Ca, the main focus of their empirical coverage figures. For comparison, we have performed the same analysis described here for the $^{48}$Ca(p,p) and the $^{208}$Pb(p,p)$^{208}$Pb cases and computed the corresponding empirical coverage. Our results from this analysis are consistent with those in \cite{King2019} (we note that these are two independent implementations of MCMC and of the underlying scattering code). The agreement is due to serendipitous cancellation between the $1/N$ factor in the likelihood and the fact that the experimental uncertainties are likely larger than the purely statistical experimental uncertainties reported. In the $^{48}$Ca example, the best-fit of the model is only able to reproduce the experimental data at a level about four times larger than the experimental data uncertainties -- that is, $\chi^{2}_{opt} \approx 16$. If the likelihood is rescaled by $1/N$, with $N=23$ for the Ca case, there is a fortuitous cancellation between the underestimated uncertainties and the inclusion of a factor of $\sqrt{N}=4.8$ in the likelihood function for MCMC, causing the MCMC-$\chi^{2}/N$ empirical coverage to lie near the diagonal. Conversely, the MCMC-$\chi^{2}$ and LM-$\chi^{2}$ results yield low empirical coverages, indicating that the experimental uncertainties may be underestimated, the model may be insufficiently flexible to reproduce data, or both. While we have no general remedy for possibly underestimated experimental uncertainty, data correlations, and a overly simply nuclear model, we hope that awareness of these challenges will clarify the limitations of our models.

\section{Conclusions}
\label{sec:conclusions}

In this study we revisit the frequentist/Bayesian comparison of \cite{King2019} considering first a simple linear problem and then add complexity to the problem to illustrate how the comparison may yield different results, including where a Bayesian treatment provides advantage for scattering problems.
Following previous work \cite{King2019}, we also consider results using a renormalized $\chi^2/N$ in the definition of the likelihood function (with $N$ being the number of data points) to account for possible correlations in the model and underestimation of the error in the data. 

Uncertainty intervals obtained in the frequentist approach (Levenberg-Marquardt algorithm for $\chi^2$ minimization) and the Bayesian approach (using Markov Chain Monte Carlo sampling to obtain parameter posterior distributions) are fully consistent for a linear model. In that case, renormalizing the $\chi^2$ by $1/N$ results in an overestimation of uncertainty by a factor of $\sqrt{N}$, which is equivalent to increasing the error on the training data by $\sqrt{N}$. 

To compare directly with recent reaction-theory analyses, we considered a simplified optical model applied to the elastic angular distributions for the scattering of neutrons off $^{90}$Zr at 10 MeV, either in a 5-parameter fit or a 6-parameter fit. For the 5-parameter fit, the frequentist and Bayesian approaches again recover the same parameter optimum and uncertainty intervals in the angular distributions, like in the linear model, despite the fact that the 5-parameter optical model is highly non-linear. For this case, the parameter posterior is well-described by a Gaussian and therefore a covariance-only frequentist approach is reliable. The same is not true for the 6-parameter analysis of neutrons off $^{90}$Zr at 10 MeV. In this case, the parameter posterior has additional non-Gaussian structure, and the covariance-based frequentist prediction becomes unreliable and a Bayesian analysis warranted.

We also considered two different definitions for the likelihood, one using the standard $\chi^2$ and another with the renormalized $\chi^2/N$. The addition of this $1/N$ factor was applied by several recent reaction-theory analyses to effect an increase of parameter uncertainty estimates, including any from correlations not accounted for in the weighted-least-squares likelihood. With this factor included, the uncertainties obtained in the cross sections increase systematically by $\sqrt{N}$. To assess the plausibility of this choice, we turn to the empirical coverages. While for the $^{90}$Zr application, the correct empirical coverages are obtained when the standard $\chi^2$ is used, for the $^{48}$Ca case discussed in \cite{King2019} the empirical coverages obtained with the standard $\chi^2$ are underestimated, a problem that is serendipitously addressed by the $1/N$ factor. 
This study demonstrates that, in general, the $1/N$ factor should not be used in the $\chi^2$. Instead, an empirical coverage test should be performed and used to assess the need for relaxing the weighted-least-squares likelihood to a more general likelihood that can account for model uncertainties, experimental error underestimation, or correlations.

\begin{acknowledgements}
We thank Dick Furnstahl, Frederi Viens, and Sofia Quaglioni for useful discussions.
 A.~E.~L. acknowledges the support of the Laboratory Directed Research and Development program of Los Alamos National Laboratory. This work was performed under the auspices of the U.S. Department of Energy by Lawrence Livermore National Laboratory under Contract No. DE-AC52-07NA27344 and by Los Alamos National Laboratory under Contract 89233218CNA000001. F. M. N. acknowledges the support of  the U.S. Department of Energy grant DE-SC0021422 and the National Science Foundation CSSI program under award number OAC-2004601 (BAND Collaboration).

\end{acknowledgements}

\bibliographystyle{apsrev}
\bibliography{Biblio}

\begin{thebibliography}{28}
\expandafter\ifx\csname natexlab\endcsname\relax\def\natexlab#1{#1}\fi
\expandafter\ifx\csname bibnamefont\endcsname\relax
  \def\bibnamefont#1{#1}\fi
\expandafter\ifx\csname bibfnamefont\endcsname\relax
  \def\bibfnamefont#1{#1}\fi
\expandafter\ifx\csname citenamefont\endcsname\relax
  \def\citenamefont#1{#1}\fi
\expandafter\ifx\csname url\endcsname\relax
  \def\url#1{\texttt{#1}}\fi
\expandafter\ifx\csname urlprefix\endcsname\relax\def\urlprefix{URL }\fi
\providecommand{\bibinfo}[2]{#2}
\providecommand{\eprint}[2][]{\url{#2}}

\bibitem[{\citenamefont{Lovell and Nunes}(2015)}]{Lovell2015}
\bibinfo{author}{\bibfnamefont{A.~E.} \bibnamefont{Lovell}} \bibnamefont{and}
  \bibinfo{author}{\bibfnamefont{F.~M.} \bibnamefont{Nunes}},
  \bibinfo{journal}{Journal of Physics G: Nuclear and Particle Physics}
  \textbf{\bibinfo{volume}{42}}, \bibinfo{pages}{034014}
  (\bibinfo{year}{2015}),
  \urlprefix\url{https://doi.org/10.1088%2F0954-3899%2F42%2F3%2F034014}.

\bibitem[{\citenamefont{Lovell et~al.}(2017)\citenamefont{Lovell, Nunes,
  Sarich, and Wild}}]{Lovell2017}
\bibinfo{author}{\bibfnamefont{A.~E.} \bibnamefont{Lovell}},
  \bibinfo{author}{\bibfnamefont{F.~M.} \bibnamefont{Nunes}},
  \bibinfo{author}{\bibfnamefont{J.}~\bibnamefont{Sarich}}, \bibnamefont{and}
  \bibinfo{author}{\bibfnamefont{S.~M.} \bibnamefont{Wild}},
  \bibinfo{journal}{Phys. Rev. C} \textbf{\bibinfo{volume}{95}},
  \bibinfo{pages}{024611} (\bibinfo{year}{2017}),
  \urlprefix\url{https://link.aps.org/doi/10.1103/PhysRevC.95.024611}.

\bibitem[{\citenamefont{King et~al.}(2018)\citenamefont{King, Lovell, and
  Nunes}}]{King2018}
\bibinfo{author}{\bibfnamefont{G.~B.} \bibnamefont{King}},
  \bibinfo{author}{\bibfnamefont{A.~E.} \bibnamefont{Lovell}},
  \bibnamefont{and} \bibinfo{author}{\bibfnamefont{F.~M.} \bibnamefont{Nunes}},
  \bibinfo{journal}{Phys. Rev. C} \textbf{\bibinfo{volume}{98}},
  \bibinfo{pages}{044623} (\bibinfo{year}{2018}).

\bibitem[{\citenamefont{King et~al.}(2019)\citenamefont{King, Lovell,
  Neufcourt, and Nunes}}]{King2019}
\bibinfo{author}{\bibfnamefont{G.~B.} \bibnamefont{King}},
  \bibinfo{author}{\bibfnamefont{A.~E.} \bibnamefont{Lovell}},
  \bibinfo{author}{\bibfnamefont{L.}~\bibnamefont{Neufcourt}},
  \bibnamefont{and} \bibinfo{author}{\bibfnamefont{F.~M.} \bibnamefont{Nunes}},
  \bibinfo{journal}{Phys. Rev. Lett.} \textbf{\bibinfo{volume}{122}},
  \bibinfo{pages}{232502} (\bibinfo{year}{2019}),
  \urlprefix\url{https://link.aps.org/doi/10.1103/PhysRevLett.122.232502}.

\bibitem[{\citenamefont{Catacora-Rios et~al.}(2019)\citenamefont{Catacora-Rios,
  King, Lovell, and Nunes}}]{Catacora2019}
\bibinfo{author}{\bibfnamefont{M.}~\bibnamefont{Catacora-Rios}},
  \bibinfo{author}{\bibfnamefont{G.~B.} \bibnamefont{King}},
  \bibinfo{author}{\bibfnamefont{A.~E.} \bibnamefont{Lovell}},
  \bibnamefont{and} \bibinfo{author}{\bibfnamefont{F.~M.} \bibnamefont{Nunes}},
  \bibinfo{journal}{Phys. Rev. C} \textbf{\bibinfo{volume}{100}},
  \bibinfo{pages}{064615} (\bibinfo{year}{2019}),
  \urlprefix\url{https://link.aps.org/doi/10.1103/PhysRevC.100.064615}.

\bibitem[{\citenamefont{Catacora-Rios et~al.}(2021)\citenamefont{Catacora-Rios,
  King, Lovell, and Nunes}}]{Catacora2021}
\bibinfo{author}{\bibfnamefont{M.}~\bibnamefont{Catacora-Rios}},
  \bibinfo{author}{\bibfnamefont{G.~B.} \bibnamefont{King}},
  \bibinfo{author}{\bibfnamefont{A.~E.} \bibnamefont{Lovell}},
  \bibnamefont{and} \bibinfo{author}{\bibfnamefont{F.~M.} \bibnamefont{Nunes}},
  \bibinfo{journal}{Phys. Rev. C} \textbf{\bibinfo{volume}{104}},
  \bibinfo{pages}{064611} (\bibinfo{year}{2021}),
  \urlprefix\url{https://link.aps.org/doi/10.1103/PhysRevC.104.064611}.

\bibitem[{\citenamefont{Whitehead et~al.}(2022)\citenamefont{Whitehead,
  Poxon-Pearson, Nunes, and Potel}}]{Whitehead2022}
\bibinfo{author}{\bibfnamefont{T.~R.} \bibnamefont{Whitehead}},
  \bibinfo{author}{\bibfnamefont{T.}~\bibnamefont{Poxon-Pearson}},
  \bibinfo{author}{\bibfnamefont{F.~M.} \bibnamefont{Nunes}}, \bibnamefont{and}
  \bibinfo{author}{\bibfnamefont{G.}~\bibnamefont{Potel}},
  \bibinfo{journal}{Phys. Rev. C} \textbf{\bibinfo{volume}{105}},
  \bibinfo{pages}{054611} (\bibinfo{year}{2022}),
  \urlprefix\url{https://link.aps.org/doi/10.1103/PhysRevC.105.054611}.

\bibitem[{\citenamefont{S\"urer et~al.}(2022)\citenamefont{S\"urer, Nunes,
  Plumlee, and Wild}}]{Surer2022}
\bibinfo{author}{\bibfnamefont{O.}~\bibnamefont{S\"urer}},
  \bibinfo{author}{\bibfnamefont{F.~M.} \bibnamefont{Nunes}},
  \bibinfo{author}{\bibfnamefont{M.}~\bibnamefont{Plumlee}}, \bibnamefont{and}
  \bibinfo{author}{\bibfnamefont{S.~M.} \bibnamefont{Wild}},
  \bibinfo{journal}{Phys. Rev. C} \textbf{\bibinfo{volume}{106}},
  \bibinfo{pages}{024607} (\bibinfo{year}{2022}),
  \urlprefix\url{https://link.aps.org/doi/10.1103/PhysRevC.106.024607}.

\bibitem[{\citenamefont{Catacora-Rios et~al.}(2023)\citenamefont{Catacora-Rios,
  Lovell, and Nunes}}]{Catacora2023}
\bibinfo{author}{\bibfnamefont{M.}~\bibnamefont{Catacora-Rios}},
  \bibinfo{author}{\bibfnamefont{A.~E.} \bibnamefont{Lovell}},
  \bibnamefont{and} \bibinfo{author}{\bibfnamefont{F.~M.} \bibnamefont{Nunes}},
  \bibinfo{journal}{Phys. Rev. C} \textbf{\bibinfo{volume}{108}},
  \bibinfo{pages}{024601} (\bibinfo{year}{2023}),
  \urlprefix\url{https://link.aps.org/doi/10.1103/PhysRevC.108.024601}.

\bibitem[{\citenamefont{Hebborn
  et~al.}(2023{\natexlab{a}})\citenamefont{Hebborn, Whitehead, Lovell, and
  Nunes}}]{Hebborn2023}
\bibinfo{author}{\bibfnamefont{C.}~\bibnamefont{Hebborn}},
  \bibinfo{author}{\bibfnamefont{T.~R.} \bibnamefont{Whitehead}},
  \bibinfo{author}{\bibfnamefont{A.~E.} \bibnamefont{Lovell}},
  \bibnamefont{and} \bibinfo{author}{\bibfnamefont{F.~M.} \bibnamefont{Nunes}},
  \bibinfo{journal}{Phys. Rev. C} \textbf{\bibinfo{volume}{108}},
  \bibinfo{pages}{014601} (\bibinfo{year}{2023}{\natexlab{a}}).

\bibitem[{\citenamefont{Hebborn
  et~al.}(2023{\natexlab{b}})\citenamefont{Hebborn, Nunes, and
  Lovell}}]{Hebborn2023PRL}
\bibinfo{author}{\bibfnamefont{C.}~\bibnamefont{Hebborn}},
  \bibinfo{author}{\bibfnamefont{F.~M.} \bibnamefont{Nunes}}, \bibnamefont{and}
  \bibinfo{author}{\bibfnamefont{A.~E.} \bibnamefont{Lovell}},
  \bibinfo{journal}{Phys. Rev. Lett.} \textbf{\bibinfo{volume}{131}},
  \bibinfo{pages}{212503} (\bibinfo{year}{2023}{\natexlab{b}}),
  \urlprefix\url{https://link.aps.org/doi/10.1103/PhysRevLett.131.212503}.

\bibitem[{\citenamefont{Trotta}(2008)}]{Trotta2008}
\bibinfo{author}{\bibfnamefont{R.}~\bibnamefont{Trotta}},
  \bibinfo{journal}{Contemporary Physics} \textbf{\bibinfo{volume}{49}},
  \bibinfo{pages}{71} (\bibinfo{year}{2008}),
  \eprint{https://doi.org/10.1080/00107510802066753},
  \urlprefix\url{https://doi.org/10.1080/00107510802066753}.

\bibitem[{\citenamefont{Thompson and Nunes}(2009)}]{ReactionsBook}
\bibinfo{author}{\bibfnamefont{I.~J.} \bibnamefont{Thompson}} \bibnamefont{and}
  \bibinfo{author}{\bibfnamefont{F.~M.} \bibnamefont{Nunes}},
  \emph{\bibinfo{title}{Nuclear Reactions for Astrophysics}}
  (\bibinfo{publisher}{Cambridge University Press}, \bibinfo{year}{2009}).

\bibitem[{\citenamefont{Neudecker et~al.}(2020)\citenamefont{Neudecker, Smith,
  Tovesson, Capote, White, Bowden, Snyder, Carlson, Casperson, Pronyaev
  et~al.}}]{Neudecker2020}
\bibinfo{author}{\bibfnamefont{D.}~\bibnamefont{Neudecker}},
  \bibinfo{author}{\bibfnamefont{D.}~\bibnamefont{Smith}},
  \bibinfo{author}{\bibfnamefont{F.}~\bibnamefont{Tovesson}},
  \bibinfo{author}{\bibfnamefont{R.}~\bibnamefont{Capote}},
  \bibinfo{author}{\bibfnamefont{M.}~\bibnamefont{White}},
  \bibinfo{author}{\bibfnamefont{N.}~\bibnamefont{Bowden}},
  \bibinfo{author}{\bibfnamefont{L.}~\bibnamefont{Snyder}},
  \bibinfo{author}{\bibfnamefont{A.}~\bibnamefont{Carlson}},
  \bibinfo{author}{\bibfnamefont{R.}~\bibnamefont{Casperson}},
  \bibinfo{author}{\bibfnamefont{V.}~\bibnamefont{Pronyaev}},
  \bibnamefont{et~al.}, \bibinfo{journal}{Nuclear Data Sheets}
  \textbf{\bibinfo{volume}{163}}, \bibinfo{pages}{228} (\bibinfo{year}{2020}),
  ISSN \bibinfo{issn}{0090-3752},
  \urlprefix\url{https://www.sciencedirect.com/science/article/pii/S0090375219300729}.

\bibitem[{\citenamefont{Thompson}(1988)}]{fresco}
\bibinfo{author}{\bibfnamefont{I.~J.} \bibnamefont{Thompson}},
  \bibinfo{journal}{Comput. Phys. Rep.} \textbf{\bibinfo{volume}{7}},
  \bibinfo{pages}{167} (\bibinfo{year}{1988}),
  \urlprefix\url{https://www.sciencedirect.com/science/article/pii/0167797788900056}.

\bibitem[{\citenamefont{Lovell}(2017)}]{quilt-r}
\bibinfo{author}{\bibfnamefont{A.~E.} \bibnamefont{Lovell}},
  \bibinfo{journal}{QUILT-R: Code on Quantification of Uncertainty in
  Low-energy Theory for Reactions}  (\bibinfo{year}{2017}).

\bibitem[{\citenamefont{Lovell et~al.}(2020)\citenamefont{Lovell, Nunes,
  Catacora-Rios, and King}}]{Lovell2021}
\bibinfo{author}{\bibfnamefont{A.~E.} \bibnamefont{Lovell}},
  \bibinfo{author}{\bibfnamefont{F.~M.} \bibnamefont{Nunes}},
  \bibinfo{author}{\bibfnamefont{M.}~\bibnamefont{Catacora-Rios}},
  \bibnamefont{and} \bibinfo{author}{\bibfnamefont{G.~B.} \bibnamefont{King}},
  \bibinfo{journal}{J. Phys. G: Nucl. Part. Phys.}
  \textbf{\bibinfo{volume}{48}}, \bibinfo{pages}{014001}
  (\bibinfo{year}{2020}).

\bibitem[{\citenamefont{Lovell and Nunes}(2018)}]{lovell2018}
\bibinfo{author}{\bibfnamefont{A.~E.} \bibnamefont{Lovell}} \bibnamefont{and}
  \bibinfo{author}{\bibfnamefont{F.~M.} \bibnamefont{Nunes}},
  \bibinfo{journal}{Phys. Rev. C} \textbf{\bibinfo{volume}{97}},
  \bibinfo{pages}{064612} (\bibinfo{year}{2018}),
  \urlprefix\url{https://link.aps.org/doi/10.1103/PhysRevC.97.064612}.

\bibitem[{\citenamefont{Levenberg}(1944)}]{Levenberg}
\bibinfo{author}{\bibfnamefont{K.}~\bibnamefont{Levenberg}},
  \bibinfo{journal}{Quarterly of Applied Mathematics}
  \textbf{\bibinfo{volume}{2}}, \bibinfo{pages}{164} (\bibinfo{year}{1944}),
  ISSN \bibinfo{issn}{0033569X, 15524485},
  \urlprefix\url{http://www.jstor.org/stable/43633451}.

\bibitem[{\citenamefont{Marquardt}(1963)}]{Marquardt}
\bibinfo{author}{\bibfnamefont{D.~W.} \bibnamefont{Marquardt}},
  \bibinfo{journal}{Journal of the Society for Industrial and Applied
  Mathematics} \textbf{\bibinfo{volume}{11}}, \bibinfo{pages}{431}
  (\bibinfo{year}{1963}), ISSN \bibinfo{issn}{03684245},
  \urlprefix\url{http://www.jstor.org/stable/2098941}.

\bibitem[{\citenamefont{Phillips et~al.}(2021)\citenamefont{Phillips,
  Furnstahl, Heinz, Maiti, Nazarewicz, Nunes, Plumlee, Pratola, Pratt, Viens
  et~al.}}]{Phillips2021}
\bibinfo{author}{\bibfnamefont{D.~R.} \bibnamefont{Phillips}},
  \bibinfo{author}{\bibfnamefont{R.~J.} \bibnamefont{Furnstahl}},
  \bibinfo{author}{\bibfnamefont{U.}~\bibnamefont{Heinz}},
  \bibinfo{author}{\bibfnamefont{T.}~\bibnamefont{Maiti}},
  \bibinfo{author}{\bibfnamefont{W.}~\bibnamefont{Nazarewicz}},
  \bibinfo{author}{\bibfnamefont{F.~M.} \bibnamefont{Nunes}},
  \bibinfo{author}{\bibfnamefont{M.}~\bibnamefont{Plumlee}},
  \bibinfo{author}{\bibfnamefont{M.~T.} \bibnamefont{Pratola}},
  \bibinfo{author}{\bibfnamefont{S.}~\bibnamefont{Pratt}},
  \bibinfo{author}{\bibfnamefont{F.~G.} \bibnamefont{Viens}},
  \bibnamefont{et~al.}, \bibinfo{journal}{Journal of Physics G: Nuclear and
  Particle Physics} \textbf{\bibinfo{volume}{48}}, \bibinfo{pages}{072001}
  (\bibinfo{year}{2021}),
  \urlprefix\url{https://dx.doi.org/10.1088/1361-6471/abf1df}.

\bibitem[{\citenamefont{Pruitt et~al.}(2023)\citenamefont{Pruitt, Escher, and
  Rahman}}]{KDUQ}
\bibinfo{author}{\bibfnamefont{C.~D.} \bibnamefont{Pruitt}},
  \bibinfo{author}{\bibfnamefont{J.~E.} \bibnamefont{Escher}},
  \bibnamefont{and} \bibinfo{author}{\bibfnamefont{R.}~\bibnamefont{Rahman}},
  \bibinfo{journal}{Phys. Rev. C} \textbf{\bibinfo{volume}{107}},
  \bibinfo{pages}{014602} (\bibinfo{year}{2023}).

\bibitem[{\citenamefont{{Neudecker, Denise}
  et~al.}(2023)\citenamefont{{Neudecker, Denise}, {Lewis, Amanda M.},
  {Matthews, Eric F.}, {Vanhoy, Jeffrey}, {Haight, Robert C.}, {Smith, Donald
  L.}, {Talou, Patrick}, {Croft, Stephen}, {Carlson, Allan D.}, {Pierson,
  Bruce} et~al.}}]{TemplateIntro}
\bibinfo{author}{\bibnamefont{{Neudecker, Denise}}},
  \bibinfo{author}{\bibnamefont{{Lewis, Amanda M.}}},
  \bibinfo{author}{\bibnamefont{{Matthews, Eric F.}}},
  \bibinfo{author}{\bibnamefont{{Vanhoy, Jeffrey}}},
  \bibinfo{author}{\bibnamefont{{Haight, Robert C.}}},
  \bibinfo{author}{\bibnamefont{{Smith, Donald L.}}},
  \bibinfo{author}{\bibnamefont{{Talou, Patrick}}},
  \bibinfo{author}{\bibnamefont{{Croft, Stephen}}},
  \bibinfo{author}{\bibnamefont{{Carlson, Allan D.}}},
  \bibinfo{author}{\bibnamefont{{Pierson, Bruce}}}, \bibnamefont{et~al.},
  \bibinfo{journal}{EPJ Nuclear Sci. Technol.} \textbf{\bibinfo{volume}{9}},
  \bibinfo{pages}{35} (\bibinfo{year}{2023}),
  \urlprefix\url{https://doi.org/10.1051/epjn/2023014}.

\bibitem[{\citenamefont{Newville et~al.}(2023)\citenamefont{Newville, Otten,
  Nelson, Stensitzki, Ingargiola, Allan, Fox, Carter, Osborn, Pustakhod
  et~al.}}]{lmfit}
\bibinfo{author}{\bibfnamefont{M.}~\bibnamefont{Newville}},
  \bibinfo{author}{\bibfnamefont{R.}~\bibnamefont{Otten}},
  \bibinfo{author}{\bibfnamefont{A.}~\bibnamefont{Nelson}},
  \bibinfo{author}{\bibfnamefont{T.}~\bibnamefont{Stensitzki}},
  \bibinfo{author}{\bibfnamefont{A.}~\bibnamefont{Ingargiola}},
  \bibinfo{author}{\bibfnamefont{D.}~\bibnamefont{Allan}},
  \bibinfo{author}{\bibfnamefont{A.}~\bibnamefont{Fox}},
  \bibinfo{author}{\bibfnamefont{F.}~\bibnamefont{Carter}},
  \bibinfo{author}{\bibfnamefont{R.}~\bibnamefont{Osborn}},
  \bibinfo{author}{\bibfnamefont{D.}~\bibnamefont{Pustakhod}},
  \bibnamefont{et~al.}, \emph{\bibinfo{title}{lmfit/lmfit-py: 1.2.2}}
  (\bibinfo{year}{2023}),
  \urlprefix\url{https://doi.org/10.5281/zenodo.8145703}.

\bibitem[{\citenamefont{{Foreman-Mackey}
  et~al.}(2013)\citenamefont{{Foreman-Mackey}, {Hogg}, {Lang}, and
  {Goodman}}}]{emcee}
\bibinfo{author}{\bibfnamefont{D.}~\bibnamefont{{Foreman-Mackey}}},
  \bibinfo{author}{\bibfnamefont{D.~W.} \bibnamefont{{Hogg}}},
  \bibinfo{author}{\bibfnamefont{D.}~\bibnamefont{{Lang}}}, \bibnamefont{and}
  \bibinfo{author}{\bibfnamefont{J.}~\bibnamefont{{Goodman}}},
  \bibinfo{journal}{Publications of the Astrophysical Society of the Pacific}
  \textbf{\bibinfo{volume}{125}}, \bibinfo{pages}{306} (\bibinfo{year}{2013}),
  \urlprefix\url{https://dx.doi.org/10.1086/670067}.

\bibitem[{\citenamefont{Wang and Rapaport}(1990)}]{Wang1990}
\bibinfo{author}{\bibfnamefont{Y.}~\bibnamefont{Wang}} \bibnamefont{and}
  \bibinfo{author}{\bibfnamefont{J.}~\bibnamefont{Rapaport}},
  \bibinfo{journal}{Nuclear Physics A} \textbf{\bibinfo{volume}{517}},
  \bibinfo{pages}{301} (\bibinfo{year}{1990}), ISSN \bibinfo{issn}{0375-9474},
  \urlprefix\url{https://www.sciencedirect.com/science/article/pii/037594749090037M}.

\bibitem[{\citenamefont{Becchetti and Greenlees}(1969)}]{BG}
\bibinfo{author}{\bibfnamefont{F.~D.} \bibnamefont{Becchetti}}
  \bibnamefont{and} \bibinfo{author}{\bibfnamefont{G.~W.}
  \bibnamefont{Greenlees}}, \bibinfo{journal}{Phys. Rev.}
  \textbf{\bibinfo{volume}{182}}, \bibinfo{pages}{1190} (\bibinfo{year}{1969}),
  \urlprefix\url{https://link.aps.org/doi/10.1103/PhysRev.182.1190}.

\bibitem[{\citenamefont{Foreman-Mackey}(2016)}]{corner}
\bibinfo{author}{\bibfnamefont{D.}~\bibnamefont{Foreman-Mackey}},
  \bibinfo{journal}{The Journal of Open Source Software}
  \textbf{\bibinfo{volume}{1}}, \bibinfo{pages}{24} (\bibinfo{year}{2016}),
  \urlprefix\url{https://doi.org/10.21105/joss.00024}.

\end{thebibliography}

\end{document}